\documentclass[journal,comsoc]{IEEEtran}
\usepackage[utf8]{inputenc}
\usepackage{xcolor}

\usepackage{graphicx}
\usepackage{booktabs}

\newcommand{\revisemajor}[1]{\textcolor{black}{#1}}
\newcommand{\reviseminor}[1]{\textcolor{black}{#1}}

\title{$4C$: A Computation, Communication, and Control Co-Design Framework for CAVs}

\author{
\IEEEauthorblockN{
Liangkai~Liu\IEEEauthorrefmark{1},
Shaoshan~Liu\IEEEauthorrefmark{2} and
Weisong~Shi\IEEEauthorrefmark{1}\\
}
\IEEEauthorblockA{
\IEEEauthorrefmark{1}Wayne State University\\
\IEEEauthorrefmark{2}PerceptIn\\
}
}

\begin{document}

\maketitle

\begin{abstract}
Connected and autonomous vehicles (CAVs) are promising due to their potential safety and efficiency benefits and have attracted massive investment and interest from government agencies, industry, and academia. 
\reviseminor{With more computing and communication resources are available, both vehicles and edge servers are equipped with a set of camera-based vision sensors, also known as Visual IoT (V-IoT) techniques, for sensing and perception.} \revisemajor{Tremendous efforts have been made for achieving programmable communication, computation, and control. However, they are conducted mainly in the silo mode, limiting the responsiveness and efficiency of handling challenging scenarios in the real world. To improve the end-to-end performance, we envision that future CAVs require the co-design of communication, computation, and control.} This paper presents our vision of the end-to-end design principle for CAVs, called $4C$, which extends the V-IoT system by providing a unified communication, computation, and control co-design framework. With programmable communications, fine-grained heterogeneous computation, and efficient vehicle controls in $4C$, CAVs can handle critical scenarios and achieve energy-efficient autonomous driving. Finally, we present several challenges to achieving the vision of the $4C$ framework.

\end{abstract}

\section{Introduction}
\label{intriduction}


The recent proliferation of computing technologies, e.g., sensors, computer vision, machine learning, and hardware acceleration, has promoted connected and autonomous vehicles (CAVs)~\cite{liu2020computing}. The global CAVs market is expected to grow to \$173.15 B by 2030, with shared mobility services contributing to 65.31\%~\cite{AVmarket}. 
In general, there are three main benefits of CAVs. The first benefit is safety. According to the fatality analysis from the National Highway Traffic Safety Administration (NHTSA), 94 percent of serious crashes are caused by human error~\cite{usdot-av-safety}. Compared with a human driver, the machine does not fatigue, drunk, and speeding. The sensors are better at distance detection, blind space detection, emergency obstacle avoidance, etc. 
The second benefit is efficiency for many aspects: less traffic congestion, less fuel consumption, less greenhouse gas emission, less travel time, etc. The improvement is because CAVs get more comprehensive traffic information, and it has better planning and controls than human drivers. The third benefit is that CAVs could support third-party applications, including applications for public safety like AMBER Alert and criminal face detection. All these applications leverage the on-vehicle computation and communication resources.

\reviseminor{CAV is one of the typical applications for V-IoT technologies~\cite{ji2019visual} since it relies on the underlying visual sensors with computation and communication devices to understand the road and context environment.} In the typical design of CAVs, sensors usually include cameras, LiDAR, radar, GPS/GNSS, etc., and computation platform includes accelerator for DNN inference, with supporting vehicle communications like dedicated short-range communications (DSRC) or cellular vehicle-to-everything (C-V2X). Besides, a real-time operating system, drivers, and complex algorithms for sensor data processing are usually implemented. \reviseminor{However, compared with other V-IoT application scenarios like smart city, the real deployment of CAVs becomes more challenging because it is safety-critical. Gigabytes of raw sensor data are generated by the visual sensors every second. How to process them in real-time with limited computation and communication resources is the key to achieving CAVs.}


\reviseminor{Lots of efforts have been made to build V-IoT systems for CAVs. For visual sensor data collection, auto-grade visual sensors like cameras, LiDAR, and radar are developed,} and the sensing range increased from less than 100 meters to over 300 meters. \reviseminor{For visual data processing,} since the perception is mainly based on deep neural networks (DNN), many accelerators have been introduced into the computation platform, including the GPU-based NVIDIA DRIVE platform, FPGA-based Zynq UltraScale+ platform, and ASIC-based Mobileye EyeQ5~\cite{grigorescu2020survey}. 
DNN compression techniques, including parameter pruning and quantization and low-rank factorization, are designed to reduce the computation demands to decrease the DNN inference time. 
The development of DSRC and C-V2X also promotes the development of autonomous driving. Through DSRC, the vehicle could read basic safety messages (BSM) from other vehicles. Besides, traffic lights, stop signs, etc., can also be shared with the vehicle from the traffic infrastructure. \revisemajor{The IEEE and 3GPP have also made efforts to achieve high reliability, low latency, and high throughput vehicle communications. The DSRC standard is evolving towards 802.11bd, while C-V2X is transitioning toward NR V2X.}
For vehicle control, vehicle drive-by-wire technology makes up the interface from the driving systems to the vehicle's control. 
\reviseminor{Recent efforts in the edge cloud system and 6G networks propose co-designing the computation, communication, and caching for task offloading algorithms~\cite{barbarossa2018edge,strinati20216g}. However, these approaches are still theoretical analyses and cannot satisfy safety-critical CAVs applications.
Therefore, current developments in computation, communication, and control are conducted either in silo mode or in theory,} limiting the responsiveness and efficiency of handling challenging real-world scenarios.



\revisemajor{Generally, with tremendous computation and communication resources equipped, V-IoT systems are deployed to both the vehicle and the edge server~\cite{ji2019visual}. The real challenge for supporting CAVs applications is coordinating the V-IoT resources to handle complex road environments, especially critical scenarios like snowing roads, heavy rain/fog, work-zone, etc. To improve the end-to-end performance for state-of-the-art driving systems,} \reviseminor{we proposed our vision of the $4C$ framework, which extends the V-IoT system by providing communication, computation, and control co-design capability for CAVs applications.} \revisemajor{4C is built on several promising technologies in communication virtualization, heterogeneous computation, and vehicle control. Unlike previous works trying to enable CAVs applications in silo mode, $4C$ is the first programming framework that provides unified APIs for managing the programmable computation, communication, and control resources. It is supposed to be deployed both on the vehicle and the edge server (V-IoT).} $4C$ is composed of three main components: programmable communications support, fine-grained heterogeneous computation, and efficient vehicle controls. Programmable communications make it possible to upload the most appropriate tasks onto the edge server (V-IoT) for computation. Fine-grained heterogeneous computation gives the programmer the ability to request heterogeneous by demand, significantly improves the flexibility. Finally, efficient vehicle controls are designed to make the feedback of the vehicle's physical control system available to the driving system. To illustrate the availability and importance of the $4C$ framework, we discuss two case studies: autonomous driving in critical scenarios, energy-efficient autonomous driving. Finally, we discuss the challenges in naming, predictability, and security. 


The rest of the paper is organized as follows. Section~\ref{motivation} presents the background and motivations of this work. Section~\ref{framework} discusses the proposed $4C$ framework. Sections~\ref{case-studies} presents two case studies of the $4C$ framework. Sections~\ref{challenges} presents the challenges and discussion. Section~\ref{conclusion} concludes the paper.

\section{Background and Motivation}
\label{motivation}


\subsection{\reviseminor{Visual IoT}}

\reviseminor{In the era of the internet of everything, billions of end devices are connected to support applications like smart cities, surveillance systems, unmanned aerial vehicles, autonomous vehicles, etc.~\cite{ji2019visual}. Visual IoT is an enabling technology which model traffic at device-level to support various applications. V-IoT is built on top of video processing technologies since rich visual sensors are deployed for environment sensing and perception ~\cite{ji2020crowd}. However, the actual deployment of V-IoT still faces several challenges. One of the biggest challenges is the visual data collection and process. Take autonomous vehicles as an example. Gigabytes of raw sensor data are generated every second and need to be processed in real-time while the vehicle is running. Therefore, the design of the visual data compression algorithm should cope with real-time requirements. Besides, the co-design of communication and computation resources would increase the V-IoT system's resource utilization. Furthermore, edge computing technology enables smart infrastructures (edge servers) equipped with sensors and computation and communication devices. More sensing and perception tasks can be executed on the edge server and share the perception results with the vehicle. Extending the current V-IoT system to support safety-critical systems becomes essential.}

\subsection{Autonomous Driving Systems}

\begin{figure}[!t]
	\centering
	\includegraphics[width=\columnwidth]{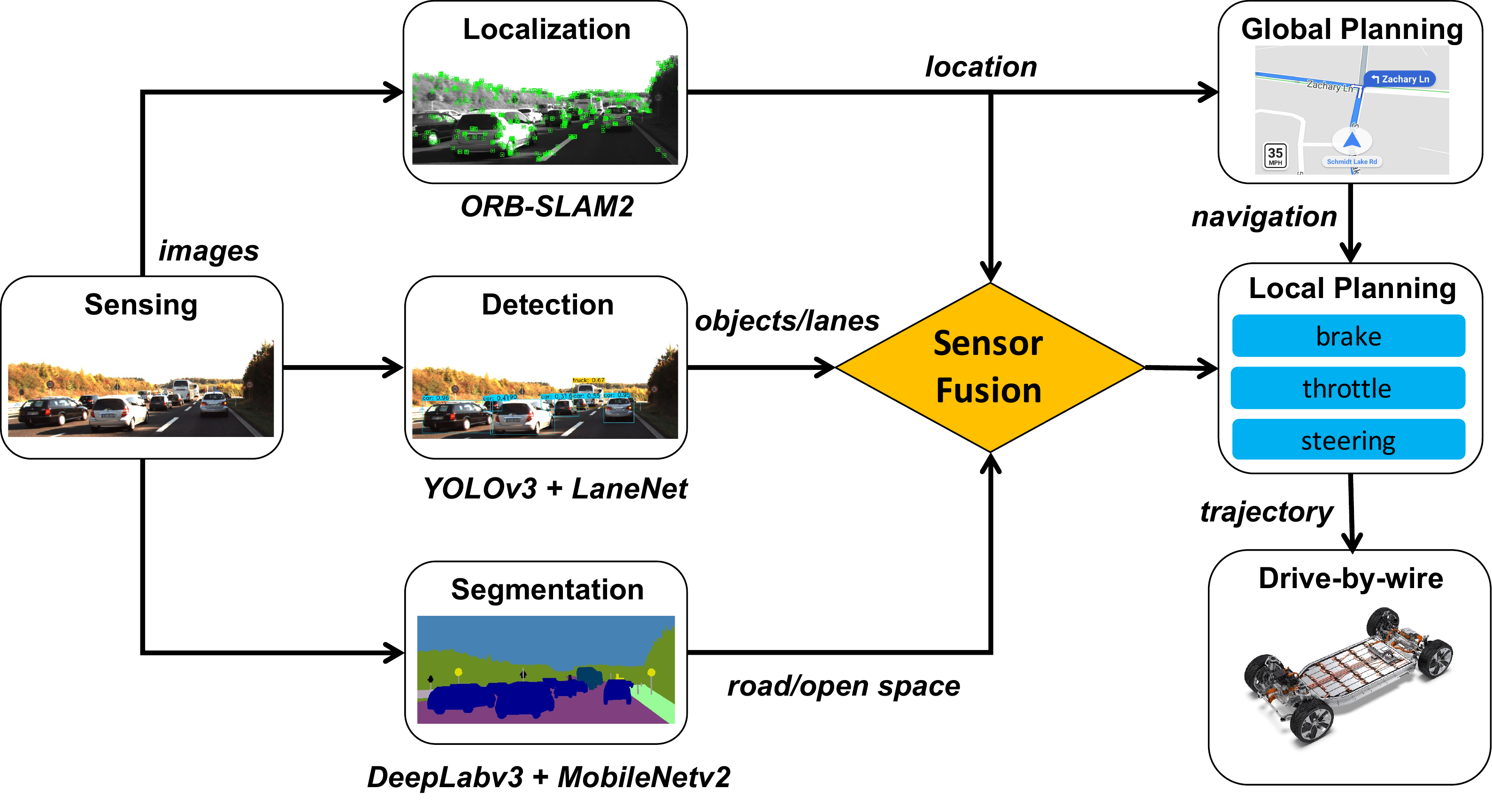}
	\caption{An overview of the end-to-end autonomous driving system.}
	\label{fig:AD-system-overview}
\end{figure}

\reviseminor{CAV is one of the typical applications for V-IoT technologies. However, unlike other application scenarios like smart cities and surveillance systems, CAV is built with dynamic wireless communications. It is a safety-critical scenario with hard real-time requirements. These characteristics make it challenging for the current V-IoT system to support CAVs applications~\cite{liu2020computing}.} Figure~\ref{fig:AD-system-overview} shows an overview of the end-to-end autonomous driving system. \revisemajor{Typically, multiple sensors are integrated to enable the vehicle's sensing and perception. To simplify the autonomous driving pipeline, we discuss a design which is purely relied on cameras for sensing the environment.} As shown in the figure, eight primary components can be divided into three parts: sensing, perception, and decision~\cite{liu2017creating}. Sensing is the process of sensors that captures information from the environment. Perception represents the understanding environment with algorithms applied to the sensing data, including localization, detection, semantic segmentation, and sensor fusion. The role of sensor fusion is to collaborate with each component's perception results and generate locations for objects, lanes, and open space for the planning module. The decision is composed of global planning, local planning, and vehicle control. Global planning generates the routes between origin and destination, while local planning generates the trajectory and controls on break, acceleration, and steering. Finally, the control commands will be applied to the drive-by-wire system and applied to the vehicle. In general, the Robot Operating System~\cite{quigley2009ros} (ROS) is used to manage these components and provide data communications between different components. 


\subsection{Why Are Vehicle Communications Important?}

Massive innovations in computer vision, machine learning, and accelerators improve the end-to-end autonomous driving systems' performance. 
However, accidents and fatalities caused by early deployed autonomous vehicles arise from time to time. There are still several challenges in the development and deployment of the autonomous driving system.

\revisemajor{
The biggest challenge is safety, which means the sensing, perception, decision, and control should be finished in real-time.} As a safety-critical system, real-time means the guarantee to finish the driving tasks before the deadline. According to~\cite{kato2015open}, when the vehicle drives 40 km per hour in urban areas and that autonomous functions should be effective every 1 m, each real-time task's execution should be less than 100ms. However, limited by DNN algorithms' performance and lack of adequate field testing, the real traffic environment is too complicated for the current driving systems to handle. 
\revisemajor{The reliability of sensors makes this problem even more challenging.} The quality of images captured by cameras is affected by the lighting conditions~\cite{liu2017creating}. Although the point clouds from LiDAR are precise and independent of illumination, there can be noisy and sparse. 
Another challenge is the cost. The cost of a level 4 autonomous driving vehicle can attain 300,000 dollars, which the sensors and computing platform cost almost 200,000 dollars~\cite{liu2020computing}. Besides, to ensure AV's reliability and safety, a backup of hardware devices may be necessary. 
The high cost of the autonomous driving vehicle remains a significant obstacle to its complete deployment in the real world.

\revisemajor{Vehicle communication mechanisms, e.g., DSRC, C-V2X, 5G, address the above challenges from another perspective. Unlike the general approaches, the vehicle communication-based approaches obtain traffic information like traffic lights, vehicle speed, pedestrians, etc., from the traffic infrastructure, decreasing computation demands for the vehicle's embedded system. Besides, with raw data updated from the vehicle, sensing and perception tasks are executed at the edge server or the cloud. In this case, co-designing the computation and communication resources for CAVs applications becomes a fundamental problem. The main reason is that as a safety-critical system, the driving system should be predictable. However, when the vehicle is driving autonomously, both the available computation and communication resources are unpredictable.} 


\subsection{Why Is Integration with Controls Important?}

\revisemajor{Typically, an essential interface for the computing system to communicate with the real vehicle is the drive-by-wire system, which provides interfaces for executing the driving commands. The drive-by-wire system sets up interactions with lower layers (vehicle system) to ensure the control commands from the upper layer (computing system) are executed correctly in real-time. However, the lack of backward interaction with the computing system makes it impossible for the upper layer to respond. Table I presents vehicle control delays for acceleration, braking, and steering for selected commercialized vehicles. Control delay is defined as the time interval between the control message's generation and execution. We can observe that the delay for any controls for all these five vehicles is larger than 100ms, while the highest exceeds 480ms. The control delay could significantly affect the vehicle's safety. Therefore, a fully autonomous driving system requires the co-design of computing and control systems.}  

\begin{table}[]
\caption{\revisemajor{Vehicle control delay for acceleration, braking, and steering for selected commercialized vehicles}}
\label{tab:control-delay}
\resizebox{\columnwidth}{!}{%
\begin{tabular}{@{}cccccc@{}}
\toprule
\textbf{Delay (ms)} &
  \textbf{\begin{tabular}[c]{@{}c@{}}Lincoln \\ MKZ\end{tabular}} &
  \textbf{\begin{tabular}[c]{@{}c@{}}Hongqi \\ H7\end{tabular}} &
  \textbf{\begin{tabular}[c]{@{}c@{}}Hongqi \\ EV\end{tabular}} &
  \textbf{\begin{tabular}[c]{@{}c@{}}NIO \\ ES8\end{tabular}} &
  \textbf{\begin{tabular}[c]{@{}c@{}}GAC Group \\ Aion LX\end{tabular}} \\ \midrule
\textit{Acceleration} & 280 & 200 & 484.2 & 120 & 236 \\
\textit{Braking}      & 230 & 362 & 191.7 & 120 & 266 \\
\textit{Steering}     & 136 & 128 & 124.8 & 108 & 120 \\ \bottomrule
\end{tabular}%
}
\end{table}

\revisemajor{There are several benefits of integrating vehicle controls with the driving system; the first is safety. The difference between the control and odometer output is road conditions like pod holes and icy roads. Under these circumstances, human drivers are supposed to decrease the speed to ensure safety, which means putting more effort into the sensing and perception stages. The second benefit is fuel efficiency. If the vehicle's control model is integrated with the path planning model, it can generate a path with the lowest fuel consumption with a safety guarantee. The third benefit is convenience, which is more related to the passenger experience for autonomous vehicles. Unlike human drivers, the machine-based system could create bang-bang controls when control switches abruptly from one extreme to the other, especially for emergency braking and obstacle avoidance. However, with V2X communications and vehicle control integrated with the driving system, a wider range of environment perception is achieved, and the vehicle controls could become smoother.}

\section{$4C$ Framework for CAVs}
\label{framework}

In the actual deployment of CAVs, how to leverage the computation, communication resources to generate and execute adequate controls under a dynamic traffic environment is a fundamental challenge. We propose the $4C$ framework, which co-designs the communication, computation, and controls to solve this challenge. In this section, we first present the overview of $4C$. Then we show a detailed discussion of computation, communication, and control, respectively. \revisemajor{Finally, we discuss the abstracted application programming interface (APIs) in 4C.}




\subsection{$4C$ Overview}

\begin{figure}[!t]
	\centering
	\includegraphics[width=\columnwidth]{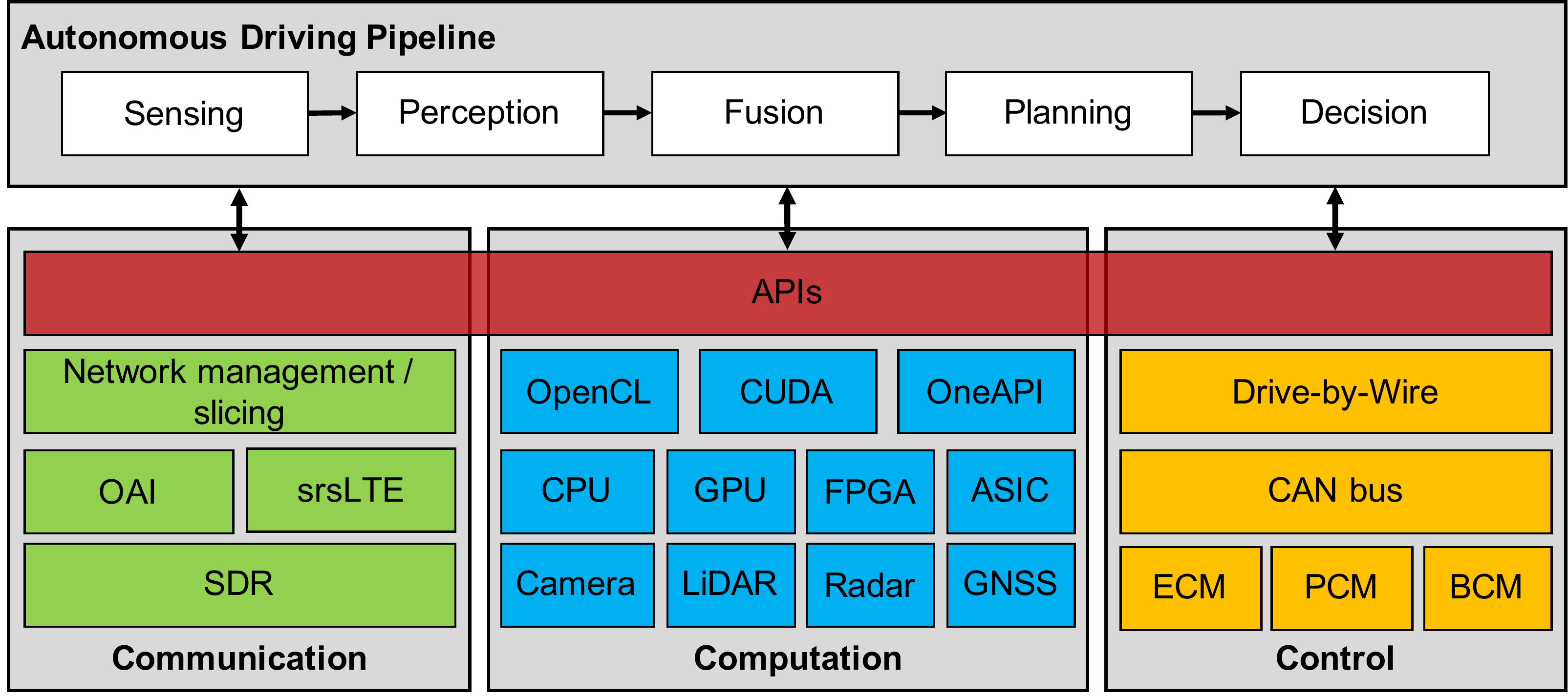}
	\caption{The overview of the $4C$ framework.}
	\label{fig:4c-overview}
\end{figure}

The overview of $4C$ is shown in Figure~\ref{fig:4c-overview}. Generally, the autonomous driving pipeline can be divided into five modules: sensing, perception, fusion, planning, and decision, where each module has several tasks running in parallel~\cite{liu2020computing}.
As a safety-critical system, the driving system requires high predictability and flexibility. To support the autonomous driving pipeline, $4C$ has three main components, including programmable communications, fine-grained heterogeneous computation, and efficient vehicle controls. \revisemajor{On top of these components, unified APIs are abstracted. $4C$ could be deployed both on the vehicle and the edge server~\cite{ji2020crowd}. The key innovation for 4C is that it provides unified APIs to manage the programmable computation, communication, and control resources.}

\subsection{Programmable Communications Support}
As a complement to the vehicle's embedded system, vehicle communications enable the vehicle to access information and computation resources with a broader range. Typically, other vehicle's speed and control information is obtained through communications. Some tasks are uploaded to the roadside unit (RSU) or cloud for computation because they have more powerful computing devices. All these capabilities require programmable communications, which could provide flexibility. Traditionally, cellular communications like 4G, LTE, and vehicle communications like DSRC are deployed in a ``black-box'' fashion, where the communication hardware and software are plug-and-play. This approach's limitation is that the devices and software cannot configure, contributing to low flexibility and programmability~\cite{bonati2020open}. Recently, many efforts have been made in pushing the softwarization and virtualization of network resources, including software-defined radio (SDR), network function virtualization (NFV), network slicing, etc.

The communication module in the $4C$ framework is built on SDR devices that have reconfiguration capability. \revisemajor{On top of that, open-source software includes OpenAirInterface (OAI), and srsLTE are implemented to provide full-stack cellular network functions~\cite{bonati2020open}.} Both OAI and srsLTE include the software stack for the core network (EPC), the base station (eNodeB), and user equipment (UE). Typically, the EPC and eNodeB are deployed on the traffic infrastructure or cloud, while the UE is deployed on vehicles. This setting makes the communication module provide 4G, LTE, 5G, and Cellular Vehicle-to-Everything (C-V2X) links between vehicles and the edge server. In the vehicle communications scenario, coverage of cellular links is limited, and lots of handovers will happen when the vehicle moves at high speed. Therefore, network management and slicing are introduced to monitor and allocate network resources at the core and base stations. Besides, context information, including the vehicle's control, the vehicle's path, the task's priority, etc., is necessary for more effective network resources management. Previously, vehicular communications only share safety-related messages like the Basic Safety Message (BSM) in DSRC. \revisemajor{In contrast, the API module in $4C$ provides reconfiguration capabilities of network resources to design CAVs applications.}

\subsection{Fine-grained Heterogeneous Computation}

\reviseminor{Visual IoT sensors are one of the primary components of the computation module, including cameras, LiDAR, Radar, GNSS, etc., which generate Gigabytes of raw sensor data every second. Processing such a huge amount of sensor data in real-time brings massive challenges for the driving systems of autonomous vehicles.} Due to its good performance and capability to process raw sensor data, DNN has been widely utilized in the driving system. How to support DNN models' execution in real-time becomes the key in driving systems design. Heterogeneous computation architecture is promising for better performance with low cost and energy budget~\cite{lin2018architectural}. However, the flexibility of current approaches to access heterogeneous computation resources is still limited, contributing to a considerable waste of computation resources. Therefore, in $4C$, we propose a fine-grained heterogeneous computation module that enables the higher-level applications to share lower-level hardware. 

From Figure~\ref{fig:4c-overview}, we can find four types of computing architectures are chosen, including CPU, GPU, Field Programmable Gate Array (FPGA), and Application Specific Integrated Circuit (ASIC)~\cite{lin2018architectural}. CPU is the most versatile architecture in computing systems, and it supports generic computations, especially good at logistic controls and sequential processing. GPU is designed mainly for graphics processing. With many cores, GPU shows strength in parallel processing like DNN model training and inference. However, the power consumption of GPU is a big concern. FPGA is an architecture based on a matrix of configurable logic blocks (CLB) connected by programmable interconnects. Therefore, FPGA shows the best reconfiguration capability, and it's much more energy-efficient than GPU. However, the programmability is much lower than the CPU and GPU. Unlike general-purpose architectures, ASIC is customized architecture for a specific logic function. This design makes ASIC shows better performance and energy efficiency than others, but the flexibility is limited. A combination of these architectures is leveraged to handle both general-purpose applications and customized functions.

Another challenge is how to access the heterogeneous hardware in fine-grained. Many efforts have been made to split the DNN model inference into several sub-tasks and execute them on the most suitable hardware. For example, the image resizing and compression can be executed on the CPU, while the matrix addition and multiplication are on GPU. Besides, the DNN model could also be split into several layers and execute on a different architecture. \revisemajor{$4C$ is designed to support both DNN layers splitting and scheduling. To program on heterogeneous architecture, $4C$ utilizes the hardware drivers in the operating system (OS) with open-source libraries, including OpenCL, CUDA, and OneAPI. On top of these libraries, $4C$ defines configuration scripts that set the destination hardware with the maximum accelerator resources to manage the splitting and scheduling of DNN layers flexibly. Similar to communication resources, APIs are defined for higher-level CAVs applications to access fine-grained heterogeneous computation.}

\subsection{Efficient Vehicle Controls}

Vehicle control is the last step of the autonomous driving pipeline, which applies decisions on the wheels. The drive-by-wire system is the bond that provides the interface to send control messages to the electronic control unit (ECU) via the controller area network (CAN) bus~\cite{isermann2002fault}. Like the I/O bus in computers, the CAN bus supports communications between different ECU and the drive-by-wire system. Typically, the vehicle is equipped with up to 80 ECUs, including engine control module (ECM), powertrain control module (PCM), Brake Control Module (BCM), etc. Each of them is an embedded system to ensure the real-time response to the commands read from the CAN bus. Take throttle control as an example; the drive-by-wire system sends a throttle command to the ECM via CAN bus, then the message will be sent to the motors to turn. Meanwhile, the throttle position sensor will monitor the throttle and send feedback to the ECM. The ECM will adjust the motor control to decrease the difference to achieve the desired throttle control.

Unlike traditional design, in the control module of the $4C$ framework, the drive-by-wire system is leveraged to bridge the driving system to ECUs like ECM, PCM, and BCM. Besides, the drive-by-wire system also reads feedback from the ECUs. The feedback of controls from the ECUs, the real-time diagnostics, and other information from the vehicle's sensor will be grouped and shared with the autonomous driving applications. The feedback of controls to the driving system sets up the stringent connection between the application and the actual vehicle, making it possible to have more customized optimizations.

\subsection{\revisemajor{APIs for Co-Design}}

\revisemajor{
From the design of the $4C$ framework, all the communication, computation, and control resources and messages are abstracted as unified APIs for the autonomous driving pipeline. This abstraction aims to enable each application to access communication and computation resources in fine-grained and customized for the specific vehicle. In summary, we can find that the APIs can be divided into two categories: management APIs and data APIs. The management APIs include managing communications bandwidth, computation resources, controls to the vehicle through the drive-by-wire system, etc. In contrast, data APIs are defined to enable each application to access runtime and context information from the lower level communication, computation, and vehicle control devices.}

\revisemajor{
We could co-design communication, computation, and control for specific applications through APIs, like energy-efficient autonomous driving, autonomous driving in critical scenarios, etc.}

\section{Case Study}
\label{case-studies}

From the design of the $4C$ framework, all the communication, computation, and control resources and messages are abstracted as unified APIs for the autonomous driving pipeline. The $4C$ framework can support various CAVs applications. This section presents two case studies of the $4C$ framework: autonomous driving in critical scenarios and energy-efficient autonomous driving.

\subsection{Autonomous Driving in Critical Scenarios}

Critical scenarios have always been a big challenge for the design and deployment of fully autonomous driving vehicles. Critical scenarios refer to the traffic environments that are extremely hard for the current driving system to understand. Currently, the field-testing of autonomous driving vehicles mostly happens in places with good weather and light conditions like Arizona and Florida. However, accidents and fatalities happen because the real traffic environment is much more complicated than the testing scenarios~\cite{liu2020computing}. Besides, DNN-based approaches are widely used in perception and planning tasks. It's tough to explain how DNN learns the pattern and knowledge as a ``black-box'' approach, contributing to many perception failure in the real traffic environment. Other critical scenarios are currently less studied, like snowing roads, heavy rain/fog, work-zone, etc. Current driving systems of autonomous vehicles are not expected to handle these scenarios. 

With the $4C$ co-design framework, some critical scenarios are addressed. The vehicle can access traffic information from other vehicles and infrastructures with the communication module, eliminating some perception failures. For critical scenarios like snowing roads, heavy rain/fog, and work-zone, the control module's feedback to the driving system could make the vehicle decelerate or emergency stop. Besides, with communications like 5G and C-V2X, real-time high-definition (HD) maps are shared with the vehicle. Furthermore, teleoperation could be launched to take over the vehicle's control for safety.

\subsection{Energy-Efficient Autonomous Driving}

With massive tasks running on the driving systems, energy consumption has been a big concern. Generally, the driving system's total power consumption exceeds 1,000 watts, and this value would be doubled if considering a duplicated system as a backup. Another major part of energy consumption is fuel/electricity consumption, mainly affected by the engine model and controls. Based on a report of the operating cost per mile for heavy-duty trucks, the fuel cost accounted for 39 percent, which is the most significant contributor to operational costs~\cite{truck-cost}. Reducing energy consumption improves the vehicle's mileage range and solving heat dissipation issues brings by overheating.

\begin{figure}[!t]
	\centering
	\includegraphics[width=\columnwidth]{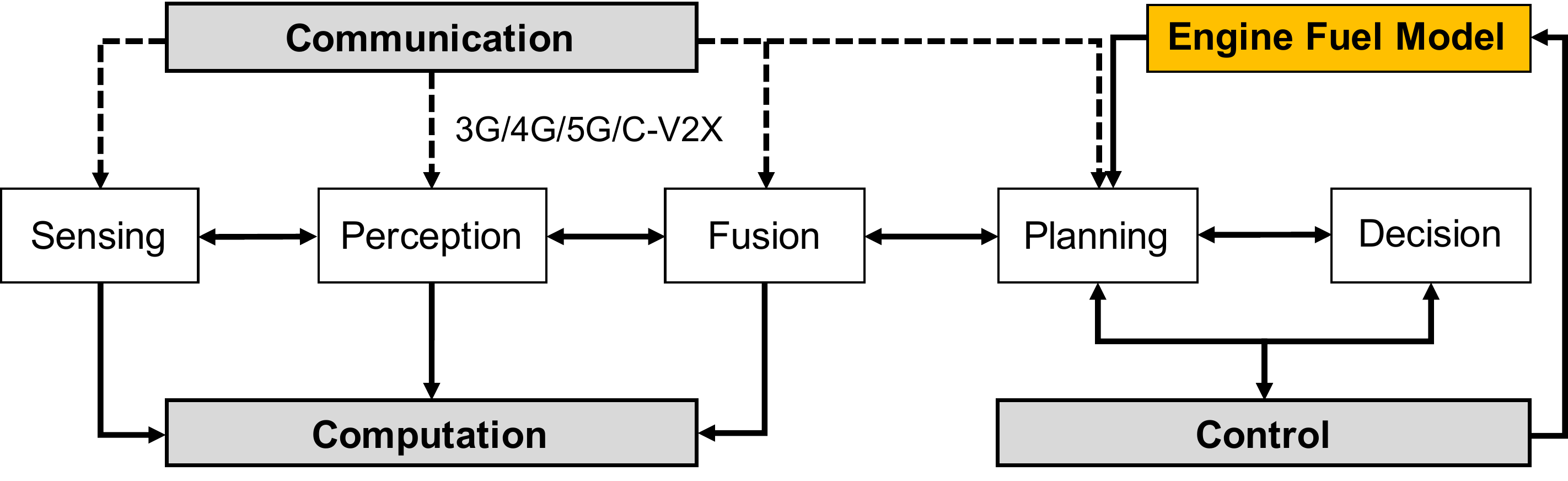}
	\caption{A typical design of energy efficient autonomous driving based on $4C$.}
	\label{fig:FEAD-case}
\end{figure}

\begin{figure}[!t]
	\centering
	\includegraphics[width=\columnwidth]{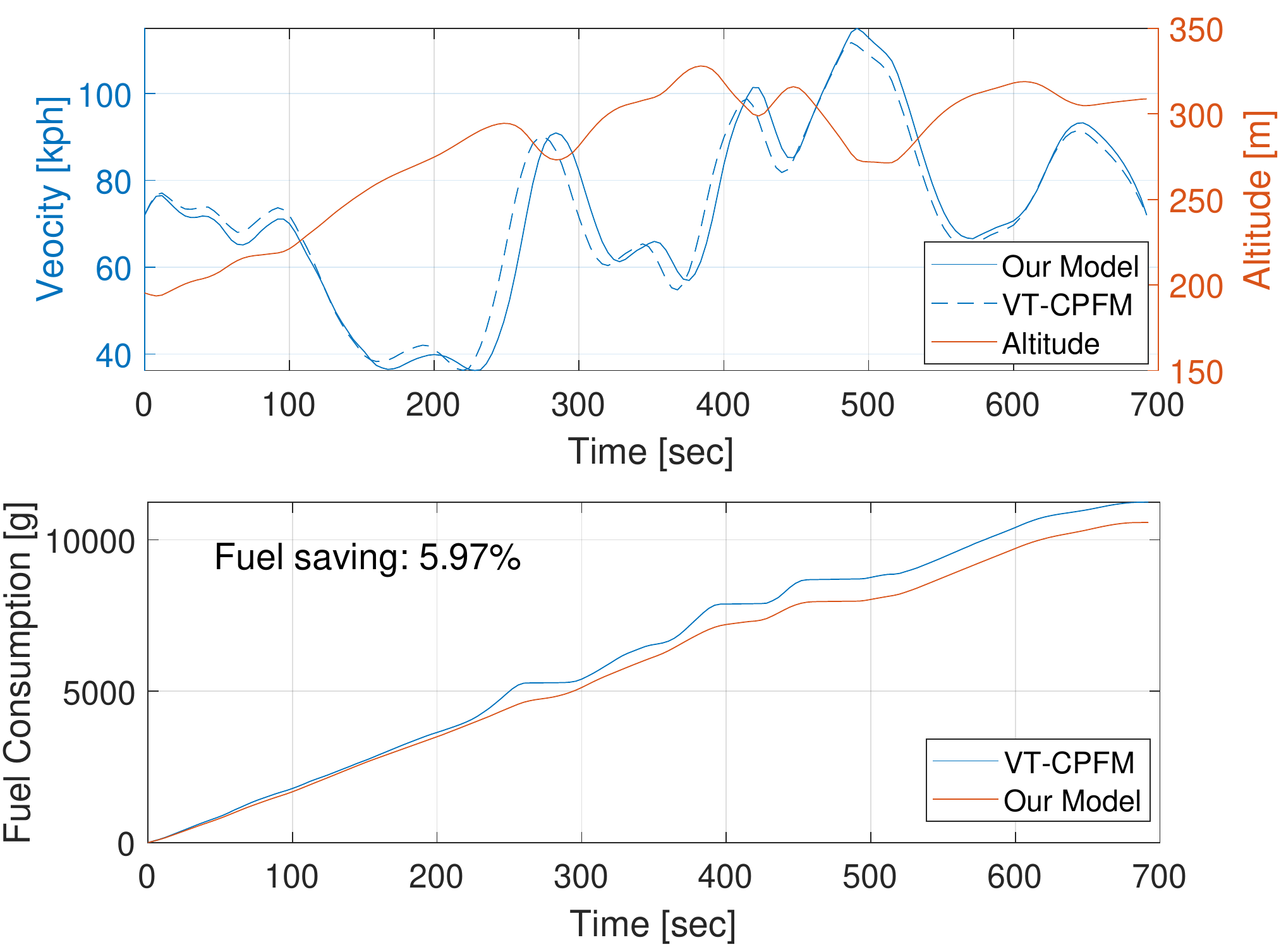}
	\caption{\revisemajor{Preliminary results of energy efficient autonomous driving by co-designing vehicle planning (computation) with vehicle controls.}}
	\label{fig:FEAD-experiments}
\end{figure}

A practical energy-efficient autonomous driving system requires the co-design of communication, computation, and control. Figure~\ref{fig:FEAD-case} shows a typical design based on the $4C$ framework. Using the recorded control data, including the throttle, brake, and torque, as the input while the corresponding fuel consumption as the label, an engine fuel model could be trained, predicting the instant and total fuel consumption for input controls. This engine fuel model could be used in path planning and the autonomous driving pipeline's decision to get a trajectory with the lowest fuel consumption. \revisemajor{To show the availability of the $4C$ framework in energy-efficient controlling, we trained a fuel rate prediction model based on recorded control data of throttle, brake, torque, etc., from the vehicle's Engine Management System (EMS). The model is trained using AutoML, which achieves a coefficient of determination ($R^2$) to 0.97. Next, we apply this fuel rate prediction model to the vehicle control module to generate the most energy-efficient control commands. We evaluate the system through a road test which is around 14 kilometers and costs 700 seconds. The comparisons of AutoML (Our Model) and state-of-the-art energy-efficient control algorithm VT-CPFM are shown in Figure~\ref{fig:FEAD-experiments}. The top subfigure shows the altitude of the road test and the optimal speed from the control module. With the optimal speed, the AutoML model and VT-CPFM model's total consumption is 10570 grams and 11240 grams, respectively, 5.97\% of total fuel saving on this road.}

In addition to energy-efficient controls, the communication module enables the vehicle to access a longer range of traffic information. The maximum range of most autonomous vehicle sensors is less than 200 meters, but the range of C-V2X communications goes to 1000 meters. Increasing the sensing range also helps the planning module to get a more energy-efficient route. Furthermore, with fine-grained access to heterogeneous computing architectures, energy-efficient scheduling becomes practical, and the power consumption of the driving system is also decreased.

\section{Challenges and Discussion}
\label{challenges}
This section presents several challenges and open problems to realize the vision of $4C$, including naming, predictability, and security. 


\textbf{Naming} 
\revisemajor{Since $4C$ is supposed to be deployed both on vehicles and the edge server}, a naming scheme becomes an essential part to support programming, addressing, and data communication between vehicles and traffic infrastructures. Unlike cellular communications, vehicle communications are ad hoc, and the duration of communication links are usually short because vehicles are driving, and the V2X communication range is limited. A traditional naming scheme like DNS cannot support this dynamic mobile scenario. In the $4C$ framework, \reviseminor{virtualized and programmable communications allow the management of V2X communications in link-level, which brings higher flexibility. However, mapping these V2X links uniquely with vehicles} with low complexity is still an open question for CAVs.

\textbf{Predictability}
An autonomous driving vehicle relies on the driving system to understand the road environment and get safe controls in real-time, requiring the whole system's predictability. The achieve the system level predictability, each stage within the autonomous driving pipeline should be predictable, which brings a massive challenge for $4C$'s communication and computation modules. \reviseminor{In 4C, with V2X communications and computations abstracted in fine-grain granularity, the behavior of tasks with link-level communications becomes more predictable. 
However, sophisticated designs are still needed to leverage abstracted resources to guarantee a real-time response.}


\textbf{Security}
As a complicated distributed system that combines computation, communication, and control, security is one of the biggest challenges for the actual deployment of CAVs applications. Currently, vehicle communication, like DSRC, is not encrypted. Attacks could happen to any hardware and software of the $4C$ framework. \reviseminor{However, with all the APIs for lower-layer resources are abstracted for higher-layer applications, plenty of mature techniques in building secure computing systems could be applied to the vehicle system.}

\section{Conclusion}
\label{conclusion}
CAVs have attracted massive attention from both the academic and automotive communities. The driving system plays a crucial role in understanding the traffic environment and making decisions for CAVs. However, the limited system-level reliability and the missing integration with vehicle communications and controls narrow the testing of CAVs to constraint scenarios. This paper presents our vision of a future driving system for CAVs, called $4C$, which provides a unified communication, computation, and control co-design framework. With two case studies, we present the usability of the $4C$ framework. Finally, we discuss the challenges in the $4C$ framework.

\ifCLASSOPTIONcaptionsoff
  \newpage
\fi

\bibliographystyle{IEEEtran}
\bibliography{main}

\begin{thebibliography}{10}
\providecommand{\url}[1]{#1}
\csname url@samestyle\endcsname
\providecommand{\newblock}{\relax}
\providecommand{\bibinfo}[2]{#2}
\providecommand{\BIBentrySTDinterwordspacing}{\spaceskip=0pt\relax}
\providecommand{\BIBentryALTinterwordstretchfactor}{4}
\providecommand{\BIBentryALTinterwordspacing}{\spaceskip=\fontdimen2\font plus
\BIBentryALTinterwordstretchfactor\fontdimen3\font minus
  \fontdimen4\font\relax}
\providecommand{\BIBforeignlanguage}[2]{{%
\expandafter\ifx\csname l@#1\endcsname\relax
\typeout{** WARNING: IEEEtran.bst: No hyphenation pattern has been}%
\typeout{** loaded for the language `#1'. Using the pattern for}%
\typeout{** the default language instead.}%
\else
\language=\csname l@#1\endcsname
\fi
#2}}
\providecommand{\BIBdecl}{\relax}
\BIBdecl

\bibitem{liu2020computing}
L.~Liu, S.~Lu, R.~Zhong, B.~Wu, Y.~Yao, Q.~Zhang, and W.~Shi, ``Computing
  systems for autonomous driving: State-of-the-art and challenges,''
  \emph{arXiv preprint arXiv:2009.14349}, 2020.

\bibitem{AVmarket}
\BIBentryALTinterwordspacing
(2018) {Global Autonomous Driving Market Outlook, 2018}. [Online]. Available:
  \url{https://www.prnewswire.com/news-releases/global-autonomous-driving-market-outlook-2018-300624588.html}
\BIBentrySTDinterwordspacing

\bibitem{usdot-av-safety}
\BIBentryALTinterwordspacing
(2020) {National Highway Traffic Safety Administration. Automated Vehicles for
  Safety.} [Online]. Available:
  \url{https://www.nhtsa.gov/technology-innovation/automatedvehicles\#topic-benefits}
\BIBentrySTDinterwordspacing

\bibitem{ji2019visual}
W.~Ji, J.~Xu, H.~Qiao, M.~Zhou, and B.~Liang, ``Visual {IoT}: Enabling internet
  of things visualization in smart cities,'' \emph{IEEE Network}, vol.~33,
  no.~2, pp. 102--110, 2019.

\bibitem{grigorescu2020survey}
S.~Grigorescu, B.~Trasnea, T.~Cocias, and G.~Macesanu, ``A survey of deep
  learning techniques for autonomous driving,'' \emph{Journal of Field
  Robotics}, vol.~37, no.~3, pp. 362--386, 2020.

\bibitem{barbarossa2018edge}
S.~Barbarossa, S.~Sardellitti, E.~Ceci, and M.~Merluzzi, ``The edge cloud: A
  holistic view of communication, computation, and caching,'' in
  \emph{Cooperative and Graph Signal Processing}.\hskip 1em plus 0.5em minus
  0.4em\relax Elsevier, 2018, pp. 419--444.

\bibitem{strinati20216g}
E.~C. Strinati and S.~Barbarossa, ``6g networks: Beyond shannon towards
  semantic and goal-oriented communications,'' \emph{Computer Networks}, vol.
  190, p. 107930, 2021.

\bibitem{ji2020crowd}
W.~Ji, B.~Liang, Y.~Wang, R.~Qiu, and Z.~Yang, ``Crowd {V-IoE}: visual internet
  of everything architecture in {AI}-driven fog computing,'' \emph{IEEE
  Wireless Communications}, vol.~27, no.~2, pp. 51--57, 2020.

\bibitem{liu2017creating}
S.~Liu, L.~Li, J.~Tang, S.~Wu, and J.-L. Gaudiot, ``Creating autonomous vehicle
  systems,'' \emph{Synthesis Lectures on Computer Science}, vol.~6, no.~1, pp.
  i--186, 2017.

\bibitem{quigley2009ros}
M.~Quigley, K.~Conley, B.~Gerkey, J.~Faust, T.~Foote, J.~Leibs, R.~Wheeler, and
  A.~Y. Ng, ``{ROS}: an open-source robot operating system,'' in \emph{ICRA
  workshop on open source software}, vol.~3, no. 3.2.\hskip 1em plus 0.5em
  minus 0.4em\relax Kobe, Japan, 2009, p.~5.

\bibitem{kato2015open}
S.~Kato, E.~Takeuchi, Y.~Ishiguro, Y.~Ninomiya, K.~Takeda, and T.~Hamada, ``An
  open approach to autonomous vehicles,'' \emph{IEEE Micro}, vol.~35, no.~6,
  pp. 60--68, 2015.

\bibitem{bonati2020open}
L.~Bonati, M.~Polese, S.~D'Oro, S.~Basagni, and T.~Melodia, ``Open,
  programmable, and virtualized 5g networks: State-of-the-art and the road
  ahead,'' \emph{arXiv preprint arXiv:2005.10027}, 2020.

\bibitem{lin2018architectural}
S.-C. Lin, Y.~Zhang, C.-H. Hsu, M.~Skach, M.~E. Haque, L.~Tang, and J.~Mars,
  ``The architectural implications of autonomous driving: Constraints and
  acceleration,'' in \emph{Proceedings of the Twenty-Third International
  Conference on Architectural Support for Programming Languages and Operating
  Systems}.\hskip 1em plus 0.5em minus 0.4em\relax ACM, 2018, pp. 751--766.

\bibitem{isermann2002fault}
R.~Isermann, R.~Schwarz, and S.~Stolzl, ``Fault-tolerant drive-by-wire
  systems,'' \emph{IEEE Control Systems Magazine}, vol.~22, no.~5, pp. 64--81,
  2002.

\bibitem{truck-cost}
\BIBentryALTinterwordspacing
(2020) {The Real Cost of Trucking – Per Mile Operating Cost of a Commercial
  Truck}. [Online]. Available:
  \url{https://www.thetruckersreport.com/infographics/cost-of-trucking/}
\BIBentrySTDinterwordspacing

\end{thebibliography}

\end{document}